\documentclass{Interspeech}

\interspeechcameraready

\title{Beyond Two-stage Diffusion TTS: Joint Structure and Content Refinement via Jump Diffusion}

\author[affiliation={1}]{Jiabao}{Ai}
\author[affiliation={1}]{Minghui}{Zhao}
\author[affiliation={1}]{Anton}{Ragni}

\affiliation{School of Computer Science}{University of Sheffield}{United Kingdom}

\email{jai5@sheffield.ac.uk, mzhao39@sheffield.ac.uk, a.ragni@sheffield.ac.uk}


\usepackage{comment}
\usepackage{booktabs}   
\usepackage{amssymb}    
\usepackage{tabularx,booktabs,makecell,amssymb}
\newcolumntype{Y}{>
{\centering\arraybackslash}X}
\usepackage{float}
\usepackage{array,tabularx,booktabs,makecell,amssymb}
\newcolumntype{Y}{>{\centering\arraybackslash}X}      
\newcolumntype{L}{>{\raggedright\arraybackslash}X}    

\usepackage{placeins}  
\usepackage{svg} 
\usepackage{algorithm}
\usepackage{algpseudocode} 
\usepackage{bm}            
\usepackage{amsmath}       
\begin{document}

\maketitle

 \begin{abstract}
    Diffusion and flow matching TTS faces a tension between discrete temporal structure and continuous spectral modeling. Two-stage models diffuse on fixed alignments, often collapsing to mean prosody; single-stage models avoid explicit durations but suffer alignment instability. We propose a jump-diffusion framework where discrete jumps model temporal structure and continuous diffusion refines spectral content within one process. Even in its one-shot degenerate form, our framework achieves 3.37\% WER vs.\ 4.38\% for Grad-TTS with improved UTMOSv2 on LJSpeech. The full iterative UDD variant further enables adaptive prosody, autonomously inserting natural pauses in out-of-distribution slow speech rather than stretching uniformly. Audio samples are available at \url{https://anonymousinterpseech.github.io/TTS_Demo/}.
\end{abstract}

\section{Introduction}

Speech is inherently dual-natured. It consists of a discrete temporal structure, namely the precise alignment and rhythmic timing of linguistic units, and continuous spectral content that realizes the acoustic details. 

In diffusion and flow-matching\cite{lipman2023flow} TTS models, alignment refers to the process of mapping input text phones or characters to temporal frames in the output speech. Alignment strategies fall into two paradigms. The first relies on explicit frame-level phone alignment, 
obtained either via learned methods such as Monotonic Alignment 
Search (MAS)~\cite{kim2020glowtts} or external forced alignment 
tools such as the Montreal Forced Aligner 
(MFA)~\cite{mcauliffe2017mfa}. Models such as Grad-TTS~\cite{popov2021gradtts}, Matcha-TTS~\cite{mehta2023matchatts} and Voicebox~\cite{le2023voicebox} first predict phone durations, upsample the sequence, and then condition diffusion on this fixed structure. This offers stability but decouples temporal structure and spectral content modeling, often leading to mean prosody and resorting to uniform mechanical stretching when adapting to total length constraints~\cite{prob_duration_tts}.

The second paradigm comprises single-stage models such as E3 TTS~\cite{gao2023e3tts}, E2 TTS~\cite{eskimez2024e2ttsembarrassinglyeasy}, F5-TTS~\cite{chen2025f5ttsfairytalerfakesfluent}, and M3-TTS~\cite{wang2025m3tts}. These forgo explicit alignment and rely on attention mechanisms. While more flexible, they suffer from instability~\cite{chen2025f5ttsfairytalerfakesfluent,zheng2025fastf5tts,wang2025m3tts}.

Although novel diffusion and flow-matching TTS models continue to appear, this fundamental dichotomy remains unresolved.
In this work, we argue that a joint generative process is essential. We introduce a jump diffusion framework~\cite{trans_dimensional} that unifies discrete jumps with continuous refinement within a single iterative process. This enables temporal structure and spectral content to co-evolve while still leveraging phone-level alignment. To reconcile variable-length jumps with standard fixed-dimensional diffusion networks, we propose an Upsample--Diffuse--Downsample (UDD) strategy.

Our main contributions are threefold: (1) a jump-diffusion framework that flexibly unifies discrete temporal structure and continuous spectral modeling---even when degraded to a one-shot variant it outperforms the regression-based Grad-TTS baseline, while the full iterative form enables adaptive prosody; (2) the UDD mechanism that enables efficient reuse of pretrained diffusion networks, circumvents the training instabilities and performance degradation observed in Trans-Dimensional-Diffusion (TDD), and supports iterative joint refinement of structure and content; and (3) empirical validation on LJSpeech showing superior intelligibility (lower WER) and natural pause insertion in out-of-distribution slow speech compared to two-stage baselines, while maintaining competitive naturalness.

\section{Background}

\subsection{Score-based Diffusion Probabilistic Modelling}
\label{ssec:score-based}
Generative modeling via diffusion\cite{sohl2015deep,ho2020denoising} can be formulated using Stochastic Differential Equations (SDEs). Following the framework established by Song et al.~\cite{song2020score}, we define a forward noising process that gradually transforms the data distribution toward a tractable reference distribution according to the Itô SDE:

\begin{equation}
    d\mathbf{x} = \mathbf{f}(\mathbf{x}, t)\,dt + g(t)\,d\mathbf{w},
\end{equation}
where $\mathbf{f}(\mathbf{x}, t)$ is the drift term, $g(t)$ is the diffusion coefficient, and $\mathbf{w}$ is a standard Wiener process. To generate data, we reverse this process from noise back to data using the corresponding reverse-time SDE, whose drift depends on the score $\nabla_{\mathbf{x}} \log p_t(\mathbf{x})$:
\begin{equation}
    d\mathbf{x} = \bigl[\mathbf{f}(\mathbf{x}, t) - g^2(t)\nabla_{\mathbf{x}} \log p_t(\mathbf{x})\bigr]\,dt + g(t)\,d\bar{\mathbf{w}}.
\end{equation}
Here, $\nabla_{\mathbf{x}} \log p_t(\mathbf{x})$ denotes the time-dependent score function, which estimates the gradient of the log-density and steers samples along the density trajectory from noise back to data. In practice, this score is approximated by a neural network $\mathbf{s}_\theta(\mathbf{x}, t)$ trained via score matching.

The same marginal density path can also be described by the deterministic probability flow ODE\cite{song2021denoising}:
\begin{equation}
    d\mathbf{x} = \left[\mathbf{f}(\mathbf{x}, t) - \tfrac12 g^2(t)\nabla_{\mathbf{x}} \log p_t(\mathbf{x})\right]dt.
\end{equation}
This formulation is particularly important because it eliminates stochastic noise while preserving the exact same marginal distributions as the reverse SDE, enabling faster and fully reproducible sampling.

\subsection{Jump Processes in Generative Modeling}

Recent work in generative modeling \cite{LIM, DLPM, GM, trans_dimensional,bertazzi2024pdgm,albergo2023stochastic} has highlighted the limitations of purely continuous diffusion and flow-matching (FM) dynamics when modeling multi-modal or heavy-tailed distributions. This is particularly relevant to TTS, where the same text can correspond to multiple valid temporal structures. Models that commit to a single fixed temporal structure may therefore collapse toward averaged timing patterns, while rare events in prosody like silent pauses remain crucial for naturalness and intelligibility \cite{DLPM, prob_duration_tts}.
A natural extension is to augment continuous diffusion with discrete jumps:
\begin{equation}
    d\mathbf{x}_t = \mathbf{f}(\mathbf{x}_t,t)\,dt + g(t)\,d\mathbf{w}_t + d\mathbf{J}_t,
\end{equation}

where $\mathbf{J}_t$ is a jump process that admits diverse designs depending on the state space and probability path~\cite{GM}; in this work, we follow the trans-dimensional formulation of~\cite{trans_dimensional}, where jumps alter the dimensionality of the state.

\section{Method}
\begin{figure}[H]
  \centering
  \includegraphics[width=\linewidth]{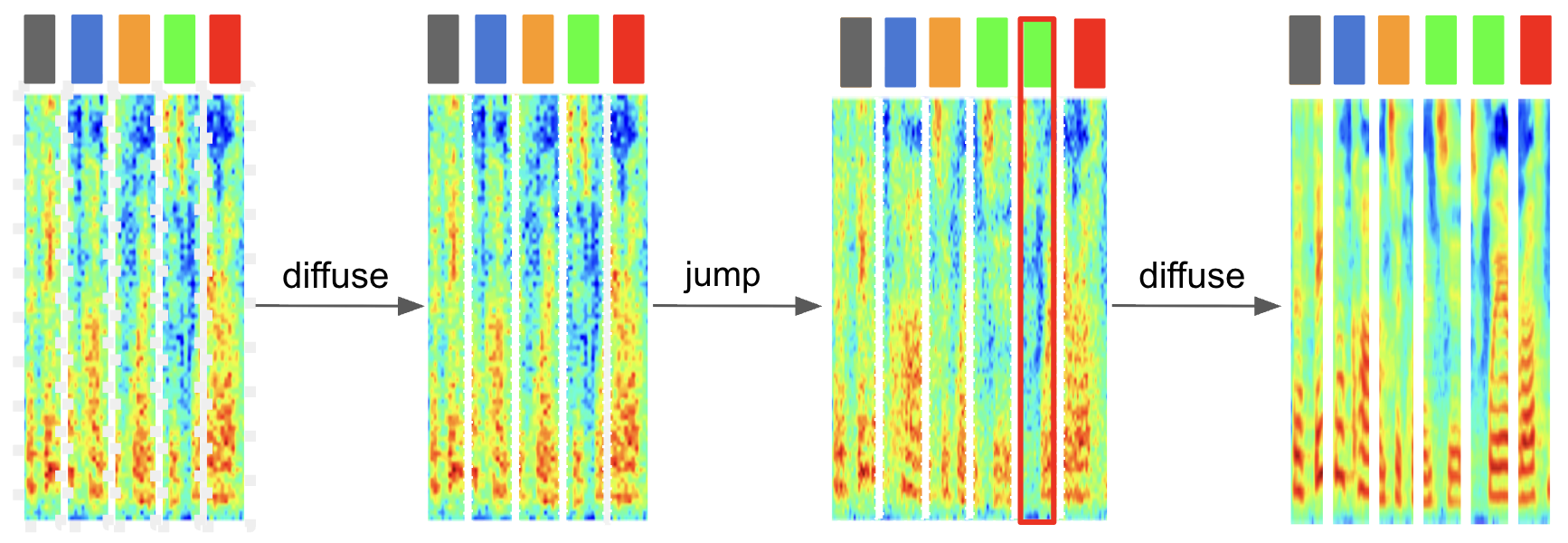}
  \caption{ jump diffusion process for Mel-spectrograms.}
  \label{fig:jump-diffusion}
\end{figure}

\begin{figure}[H]
  \centering
  \includegraphics[width=\linewidth]{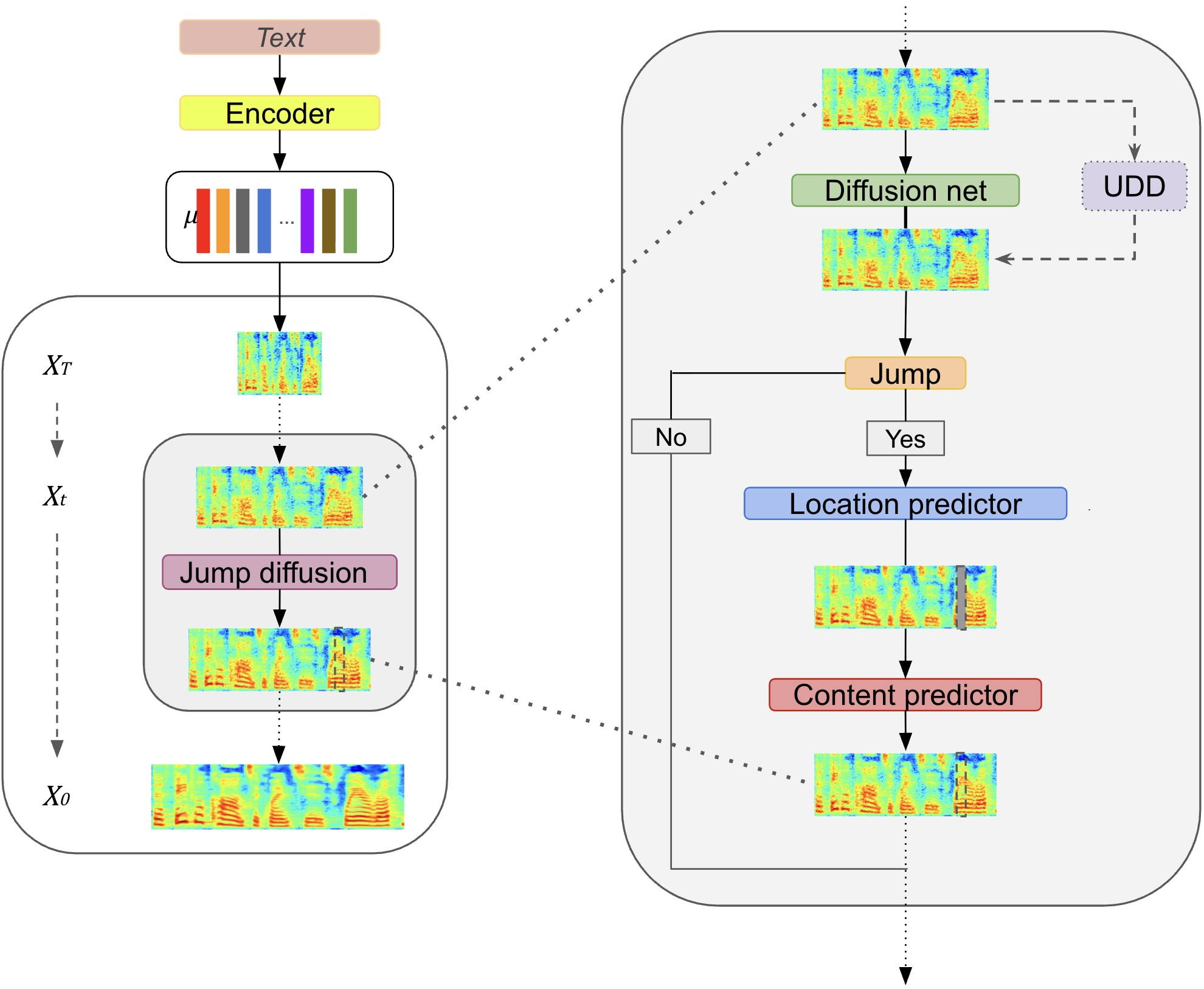}
  \caption{Inference pipeline. Starting from a noisy and incomplete phone-level state, the model iteratively performs jumps and diffusion.}
  \label{fig:inference}
\end{figure}

 We adapt this idea to TTS (Fig.~\ref{fig:jump-diffusion}, Fig.~\ref{fig:inference}): a Location Predictor determines where to insert new temporal frames, a Content Predictor generates what to fill in, and standard diffusion steps refine the spectral content.

\FloatBarrier

\subsection{Forward Jump Diffusion}
\label{ssec:forward}
The forward process \(q(\mathbf{x}_t \mid \mathbf{x}_0)\) transforms the ground-truth Mel-spectrogram \(\mathbf{x}_0 \in \mathbb{R}^{D \times L_0}\) (where \(D=80\) is the number of Mel-frequency bins and \(L_0\) is the original number of time frames) into a variable-length latent state \(\mathbf{x}_t \in \mathbb{R}^{D \times L_t}\) by coupling \textit{structural corruption} (frame deletion) and \textit{spectral corruption} (Gaussian noise addition, following the standard diffusion process in Section~\ref{ssec:score-based}).

\textbf{Structural Corruption.} We define a protected set \(\mathcal{P}\) containing the first frame of each phone (derived from forced alignment \(\mathcal{A}\)). For \(t \in [0,1]\), the target length follows the linear schedule
\begin{equation}
L_t = |\mathcal{P}| + \left\lfloor \left(1 - \frac{t - t_{\min}}{1 - t_{\min}}\right)(L_0 - |\mathcal{P}|) \right\rfloor.
\label{eq:length-schedule}
\end{equation}
Following \cite{trans_dimensional}, we set $t_{\min}=0.1$.
Frames not in \(\mathcal{P}\) are randomly removed until the target length \(L_t\) is reached, yielding the downsampled sequence \(\mathbf{x}_t^{\rm sub} \in \mathbb{R}^{D \times L_t}\) and the corresponding encoder output \(\boldsymbol{\mu}_t^{\rm sub}\).

\textbf{Spectral Corruption.} Following the variance-preserving diffusion SDE described in the Background (Eqs.~(1)--(3)), Gaussian noise is added to \(\mathbf{x}_t^{\rm sub}\) relative to the phone-level encoder output \(\boldsymbol{\mu}_t^{\rm sub}\):
\begin{align}
\mathbf{x}_t &= \mathbf{x}_t^{\rm sub} \exp\left(-\tfrac12 \int_0^t \beta_s\,ds\right) \nonumber \\
&\quad + \boldsymbol{\mu}_t^{\rm sub}\left(1 - \exp\left(-\tfrac12 \int_0^t \beta_s\,ds\right)\right) \nonumber \\
&\quad + \sqrt{1 - \exp\left(-\int_0^t \beta_s\,ds\right)} \cdot \mathbf{z}_t,
\label{eq:forward_noise}
\end{align}
where \(\mathbf{z}_t \sim \mathcal{N}(\mathbf{0}, \mathbf{I}_{D \times L_t})\).

To generate training targets for the jump predictors, we additionally simulate a single-step transition by randomly removing one column \(k\) from \(\mathbf{x}_t\), forming the triplet \((\mathbf{x}_t^{\setminus k}, s_{\rm target}, \mathbf{x}_0^k)\).

\begin{algorithm}[H]
\caption{Forward Process \& Training Target Generation}
\label{alg:forward}
\begin{algorithmic}[1]
\State \textbf{Input:} \(\mathbf{x}_0, \boldsymbol{\mu}, \mathcal{A}\); \textbf{Sample} \(t \sim \mathcal{U}(0,1)\)
\State \Comment{1. Structural Corruption}
\If{\(t > t_{\min}\)}
    \State \(L_0 \gets \text{Length}(\mathbf{x}_0)\)
    \State \(\mathcal{P} \gets\) first-frame indices of each phone in \(\mathcal{A}\)
    \State \(L_t \gets\) ScheduleLength(\(L_0, |\mathcal{P}|, t\)) \Comment{Eq.~\eqref{eq:length-schedule}}
    \State Sample \(\mathcal{I}_{\rm keep}\) of size \(L_t\) s.t. \(\mathcal{P} \subseteq \mathcal{I}_{\rm keep}\)
    \State \(\mathbf{x}_t^{\rm sub} \gets \mathbf{x}_0[:, \mathcal{I}_{\rm keep}]\), \(\boldsymbol{\mu}_t^{\rm sub} \gets \boldsymbol{\mu}[:, \mathcal{I}_{\rm keep}]\)
\Else
    \State \(\mathbf{x}_t^{\rm sub} \gets \mathbf{x}_0\), \(\boldsymbol{\mu}_t^{\rm sub} \gets \boldsymbol{\mu}\)
\EndIf
\State \Comment{2. Spectral Corruption}
\State \(\mathbf{z}_t \sim \mathcal{N}(\mathbf{0}, \mathbf{I})\)
\State \(\mathbf{x}_t \gets\) AddNoise(\(\mathbf{x}_t^{\rm sub}, \boldsymbol{\mu}_t^{\rm sub}, t, \mathbf{z}_t\))
\State \Comment{3. Single-step Jump Targets}
\State \(k \sim \mathcal{U}\{1,\dots,\text{Length}(\mathbf{x}_t)\}\)
\State \(\mathbf{x}_t^{\setminus k} \gets\) Delete(\(\mathbf{x}_t, k\)), \(\mathbf{x}_0^k \gets \mathbf{x}_t^{\rm sub}[:,k]\), \(s_{\rm target} \gets k\)
\State \Return \(\mathbf{x}_t\), \(\{\mathbf{x}_t^{\setminus k}, \mathbf{x}_0^k, s_{\rm target}\}\)
\end{algorithmic}
\end{algorithm}

\subsection{Reverse Process (Inference)}
\label{ssec:reverse}
\FloatBarrier

The reverse process starts from a compressed phone-level state \(\mathbf{x}_1 \approx \tilde{\boldsymbol{\mu}}\) (where \(\tilde{\boldsymbol{\mu}} \in \mathbb{R}^{D \times N_{\rm phone}}\) is the non-upsampled output of the frozen text encoder) and progressively grows it to the target length \(L_{\rm target}\). The target length \(L_t\) at each reverse step follows the same linear schedule as the forward process (Eq.~\eqref{eq:length-schedule}), traversed in reverse from \(L_1 = |\mathcal{P}|\) to \(L_0 = L_{\rm target}\). It interleaves discrete structural expansions with continuous spectral refinements (see Algorithm~\ref{alg:inference}).

\textbf{Structural Jump.} At step \(t\), we compute the number of insertions needed \(N_{\rm ins} = \max(0, L_t - \text{len}(\mathbf{x}_t))\). The Location Predictor (Section~\ref{ssec:implementation}) samples insertion slots \(S\), and the Content Predictor generates new frame content \(\hat{\mathbf{x}}_0\). Inserted columns in the prior \(\boldsymbol{\mu}_t\) are created by duplicating the left neighboring frame.

\textbf{Spectral Diffusion.} The expanded sequence is updated via a standard reverse diffusion step (SDE or ODE solver, Eq.~(2) in Background) to reduce spectral noise.

\begin{algorithm}[H]
\caption{Inference}
\label{alg:inference}
\begin{algorithmic}[1]
\State \(\tilde{\boldsymbol{\mu}}, L_{\rm target} \gets\) Encoder(Text)
\State \(\mathbf{x}_1 \gets \tilde{\boldsymbol{\mu}} + \mathbf{z}\); \(\boldsymbol{\mu}_t \gets \tilde{\boldsymbol{\mu}}\)
\For{\(t = 1 \to 0\)}
    \State \(N_{\rm ins} \gets \max(0, L_t - \text{len}(\mathbf{x}_t))\)
    \If{\(N_{\rm ins} > 0\)}
        \State \(S \gets\) Sample \(N_{\rm ins}\) slots from LocationNet(\(\mathbf{x}_t, \boldsymbol{\mu}_t, t\))
        \State \(\hat{\mathbf{x}}_0 \gets\) ContentNet(\(\mathbf{x}_t, \boldsymbol{\mu}_t, t, S\))
        \State \(\mathbf{x}_t \gets\) Insert(\(\mathbf{x}_t\), Noised(\(\hat{\mathbf{x}}_0, t\)), \(S\))
        \State \(\boldsymbol{\mu}_t \gets\) Insert(\(\boldsymbol{\mu}_t, \boldsymbol{\mu}_t[S], S\))
    \EndIf
    \State \(\mathbf{x}_t \gets\) DenoiseStep(\(\mathbf{x}_t, \boldsymbol{\mu}_t, t\))
\EndFor
\State \Return \(\mathbf{x}_0\)
\end{algorithmic}
\end{algorithm}

\subsection{Implementation Details}
\label{ssec:implementation}
We adopt the text encoder and U-Net diffusion backbone from the official Grad-TTS checkpoint~\cite{popov2021gradtts} and keep them frozen during training of the jump predictors. The framework introduces two additional networks:

\textbf{Content Predictor.} To reconstruct deleted frames, we use a bidirectional Transformer encoder (8 layers, 8 heads). The input is the reduced prior \(\boldsymbol{\mu}_t\) and \(\mathbf{x}_t\) with the target column zeroed (in-place masking). A 1D convolution extracts local features, followed by the Transformer for global context. The network predicts a residual \(\Delta\):
\begin{equation}
\hat{\mathbf{x}}_0^k = \boldsymbol{\mu}_t^k + \Delta.
\label{eq:content_pred}
\end{equation}
The loss is
\begin{equation}
\mathcal{L}_{\rm cont} = \lVert \hat{\mathbf{x}}_0^k - \mathbf{x}_0^k \rVert_1 + \lambda_{\rm prior} \lVert \hat{\mathbf{x}}_0^k - \boldsymbol{\mu}_t^k \rVert_2^2.
\end{equation}

\textbf{Location Predictor ($\mathcal{L}_{\rm loc}$):}
The Location Predictor identifies the optimal insertion slot \(s \in \{0, \dots, L_t-1\}\) for the generated frame, where \(L_t\) is the current sequence length at time \(t\). It utilizes a 4-layer Transformer encoder to generate frame-level representations \(\mathbf{h}_s\). A linear scoring head is applied to \(\mathbf{h}_s\) to produce logits \(o_j\). The training objective is the cross-entropy loss over the \(L_t\) possible insertion positions:
\begin{equation}
\mathcal{L}_{\rm loc} = -\sum_{j=0}^{L_t-1} y_j \log \left( \frac{\exp(o_j)}{\sum_{k=0}^{L_t-1} \exp(o_k)} \right),
\end{equation}
where \(y_j\) is the one-hot target indicating the original deletion index (i.e., the position of the frame that was removed during the forward process).

\subsection{UDD and One-shot Variant}
\label{sec:udd}
\begin{figure}[H]
  \centering
  \includegraphics[width=\linewidth]{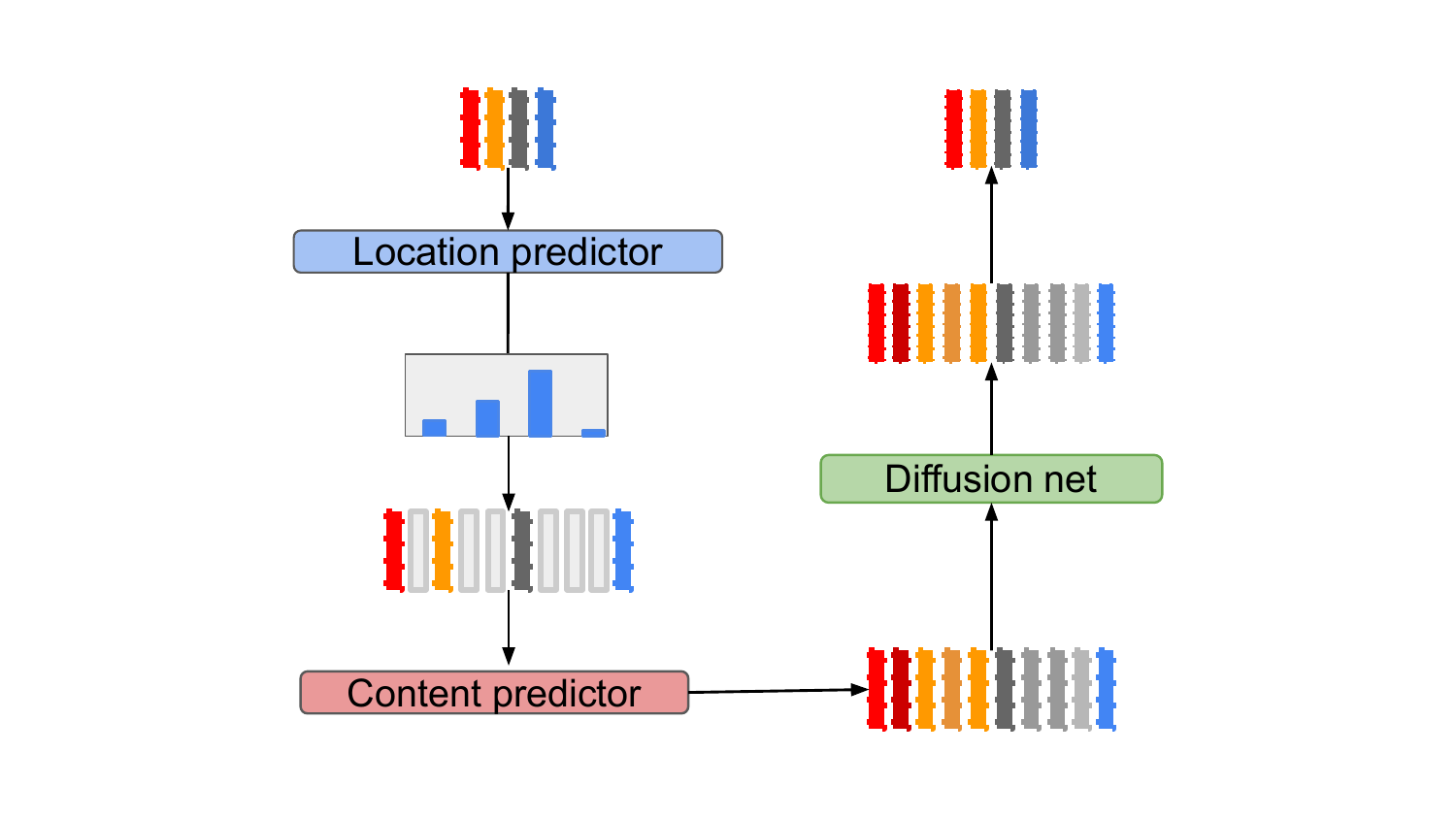}
  \caption{Upsample--Diffuse--Downsample (UDD).}
  \label{fig:udd}
\end{figure}

The Location Predictor can also act as an  upsampler. Given a sequence $\mathbf{x}_t \in \mathbb{R}^{D \times L_t}$, it predicts a distribution over $L_t$ insertion slots. To reach the target length $L_{\rm target}$, we allocate $N_{\mathrm{add}}=L_{\rm target}-L_t$ new frames according to this distribution. 

If all insertions happen in a single step at the very beginning, our model degrades to a One-shot variant in which the Location Predictor replaces Grad-TTS's regression-based duration predictor with a classification-based alternative.

This insight leads to the Upsample--Diffuse--Downsample (UDD) strategy (Fig.~\ref{fig:udd}). We first expand the variable-length state to a fixed-length canvas by inserting frames predicted by the Location and Content predictors. A standard reverse diffusion step is then applied to this full-length representation. Finally, we downsample by retaining only the original columns and proceed to the next iteration. UDD thus enables iterative joint refinement of temporal structure and spectral content while reusing the pretrained fixed-dimensional diffusion network without retraining.

\section{Experiments}
\subsection{Experimental Setup}
Experiments are conducted on the LJSpeech~\cite{ljspeech17} using the standard partitions from~\cite{popov2021gradtts}. The diffusion network and text encoder follow the Grad-TTS architecture~\cite{popov2021gradtts} and are initialized with pretrained weights. The proposed jump predictors are optimized using Adam ($lr=1\times10^{-4}$). All training hyperparameters exactly follow the original Grad-TTS recipe~\cite{popov2021gradtts}. Waveform synthesis uses a pretrained HiFi-GAN vocoder~\cite{kong2020hifigan}. For evaluation, intelligibility (WER) is measured with Whisper medium~\cite{radford2023whisper} and naturalness with UTMOSv2~\cite{baba2024utmosv2}.

\subsection{Main Results and Comparisons}
\label{ssec:main_results}

We evaluate spectral accuracy (MCD), prosodic consistency (Log-F0 RMSE), naturalness (UTMOSv2), and intelligibility (WER). To ensure a fair comparison, the total target length $L_{\rm target}$ for each utterance is kept identical to Grad-TTS across all systems.

\begin{table}[H]
\centering
\scriptsize
\setlength{\tabcolsep}{2.5pt}
\begin{tabularx}{\columnwidth}{l c c c c}
\toprule
\textbf{Method} & \textbf{WER (\%)} $\downarrow$ & \textbf{MCD} $\downarrow$ & \textbf{Log-F0 RMSE} $\downarrow$ & \textbf{UTMOSv2} $\uparrow$ \\
\midrule
Grad-TTS & 4.38 & 5.872 $\pm$ 0.52 & 0.330 $\pm$ 0.09 & 4.024 $\pm$ 0.21 \\
One-shot & \textbf{3.37} & 5.914 $\pm$ 0.55 & 0.332 $\pm$ 0.08 & \textbf{4.050} $\pm$ 0.21 \\
\midrule
TDD (Sample) & 6.31 & 6.057 $\pm$ 0.58 & 0.341 $\pm$ 0.09 & 3.867 $\pm$ 0.21 \\
TDD (Argmax) & 7.66 & 5.850 $\pm$ 0.61 & 0.341 $\pm$ 0.10 & 3.988 $\pm$ 0.25 \\
UDD (Sample) & 4.55 & 5.860 $\pm$ 0.54 & 0.344 $\pm$ 0.09 & 3.989 $\pm$ 0.20 \\
UDD (Argmax) & 4.71 & \textbf{5.830} $\pm$ 0.55 & \textbf{0.332} $\pm$ 0.08 & 4.003 $\pm$ 0.22 \\
\bottomrule
\end{tabularx}
\caption{Main evaluation results.}
\label{tab:main_results}
\end{table}

\textbf{Classification vs.\ regression for duration modeling.}
The One-shot model achieves the lowest WER (3.37\%) and highest UTMOSv2 (4.050), improving substantially over Grad-TTS (4.38\% WER, 4.024 UTMOSv2). Since One-shot replaces only the regression-based duration predictor with our classification-based Location Predictor while keeping all other components identical, this result directly confirms that modeling phone durations as a categorical distribution over insertion slots captures the multi-modal nature of speech timing better than MSE regression, which collapses toward mean durations and underrepresents rare durational events such as pauses.

\textbf{Why TDD fails.}
TDD applies the U-Net architecture directly, resulting in the worst WER across all variants (6.31--7.66\%). We attribute this to two factors. First, the input dimensionality changes at every diffusion step, drastically increasing the task uncertainty faced by the network and making learning harder. Second, the convolutional backbone possesses translation invariance but lacks scale invariance; since the sequence dimensionality changes during the process. Together, these issues cause TDD to degrade severely compared to fixed-length alternatives.

\textbf{UDD: bridging variable-length jumps and fixed-dimensional diffusion.}
UDD resolves the dimensional mismatch by temporarily upsampling the sequence to the target length before each diffusion step. This preserves the pretrained U-Net's inductive bias and yields the best MCD (5.830) and competitive Log-F0 RMSE (0.332) among all diffusion-based variants.

\subsection{Adaptive Pauses in Slow Speech}
\label{ssec:slow_speech}

The preceding experiment evaluates all systems at matched target lengths derived from ground-truth durations. We now test a more challenging scenario: synthesizing speech at $0.75\times$ speed. This out-of-distribution setting reveals how each model distributes the additional frames.

\begin{table}[H]
\centering
\scriptsize
\setlength{\tabcolsep}{4pt}
\begin{tabularx}{\columnwidth}{l c c c c}
\toprule
\textbf{Method ($0.75\times$)} & \textbf{Avg Total (s)} & \textbf{Avg Silence (s)} & \textbf{Sil.\ Ratio $\uparrow$} & \textbf{WER (\%) $\downarrow$} \\
\midrule
GT ($1.00\times$) & 6.26 & 0.45 & 7.19\% & 1.94 \\
Grad-TTS & 7.37 & 0.47 & 6.38\% & 4.29 \\
\midrule
UDD (Argmax) & 7.37 & 0.71 & \textbf{9.63\%} & \textbf{4.09} \\
UDD (Sample) & 7.37 & 0.61 & 8.28\% & 4.19 \\
\bottomrule
\end{tabularx}
\caption{Duration breakdown and intelligibility at $0.75\times$ speed. Silence Ratio = Silence\,/\,Total.}
\label{tab:slow_speed_results}
\end{table}

\begin{figure}[H]
  \centering
  \includegraphics[width=\linewidth]{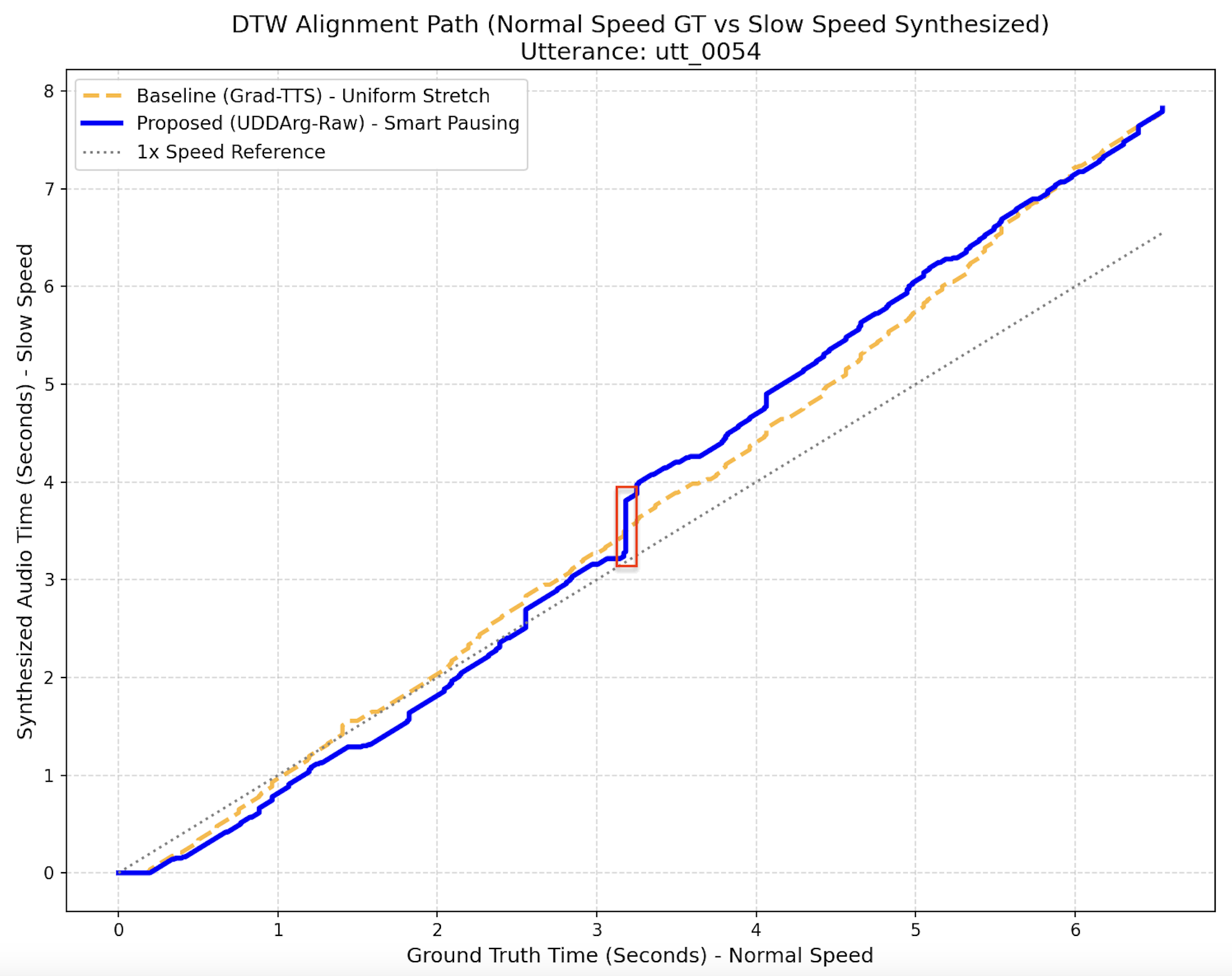}
  \caption{DTW alignment paths at $0.75\times$ speed.}
  \label{fig:dtw_alignment}
\end{figure}

\textbf{Grad-TTS stretches uniformly.} As shown in Fig.~\ref{fig:dtw_alignment}, Grad-TTS produces a near-linear DTW alignment path, indicating that the extra frames are distributed proportionally across all phones. Its silence ratio (6.38\%) is even lower than the ground truth at normal speed (7.19\%), confirming that uniform stretching dilutes pauses rather than preserving them.

\begin{figure}[H]
  \centering
  \includegraphics[width=\linewidth]{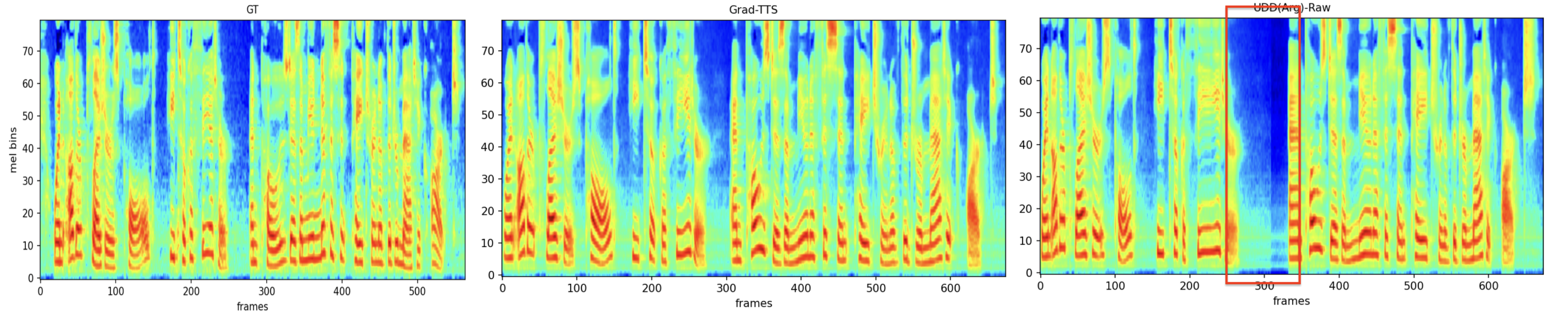}
  \caption{Example Mel-spectrograms at $0.75\times$ speed.}
  \label{fig:mel_comparison}
\end{figure}

\textbf{Jump-based models redistribute duration adaptively.} UDD allocates substantially more duration to silent pauses, with UDD (Argmax) reaching a silence ratio of 9.63\%---well above the 6.38\% of Grad-TTS and even exceeding the ground-truth level (7.19\%). Its DTW path (Fig.~\ref{fig:dtw_alignment}) and corresponding Mel-spectrograms (Fig.~\ref{fig:mel_comparison}) exhibit a clear staircase pattern, where vertical jumps correspond to the autonomous insertion of linguistically appropriate pauses. This adaptive redistribution yields the lowest WER in this setting (4.09\% vs.\ 4.29\% for Grad-TTS), confirming allocating extra duration to pauses rather than uniformly stretching all frames benefits intelligibility.

\section{Conclusion and Future Work}

We presented a jump-diffusion framework that unifies discrete temporal structure modeling and continuous spectral refinement within a single probabilistic process for TTS. Even in its simplest one-shot degenerate form, our framework achieves 3.37\% WER and 4.050 UTMOSv2, outperforming the regression-based duration modeling baseline (4.38\% WER, 4.024 UTMOSv2), confirming the advantage of classification-based duration modeling. The full iterative UDD variant further enables adaptive prosody: in slow speech, it autonomously inserts linguistically appropriate pauses rather than uniformly stretching all frames, achieving 4.09\% WER compared to 4.29\% for Grad-TTS with a silence ratio of 9.63\% versus 6.38\%.

Currently, jumps operate exclusively on temporal structure, leaving spectral content to be refined solely by continuous diffusion. Future work will explore jumps in the spectral domain to enable discrete corrections of spectral content alongside structural changes\cite{LIM, DLPM, GM, bertazzi2024pdgm}. We also plan to evaluate on multi-speaker and more spontaneous speech corpora, where the greater prosodic variability is expected to further highlight the advantage of our jump-based duration modeling~\cite{prob_duration_tts}.

\bibliographystyle{IEEEtran}
\bibliography{mybib}

\end{document}